\documentclass{myarticle}

\usepackage{txfonts}
\usepackage[colorlinks]{hyperref}
\usepackage{tikz}

\usetikzlibrary{positioning}

\newcommand{\n}{\noindent}
\newtheorem{thm}{Theorem}
\newtheorem{Proposition}{Proposition}
\newtheorem{lemma}{Lemma}
\newtheorem{cor}{Corollary}

\newtheorem{ex}{Example}

\begin{document}

\title{NARUMI-KATAYAMA INDEX OF TOTAL TRANSFORMATION GRAPHS}

\author{NILANJAN DE}
\ead{de.nilanjan@rediffmail.com}

\address{Department of
Basic Sciences and Humanities (Mathematics),\\ Calcutta Institute of Engineering and Management, Kolkata, India.}
\cortext[cor1]{Corresponding Author.}

\begin{abstract}
The Narumi-Katayama index of a graph was introduced in 1984 for representing the carbon skeleton of a saturated hydrocarbons and is defined as the product of degrees of all the vertices of the graph. In this paper, we examine the Narumi-Katayama index of  different total transformation graphs.
\end{abstract}

\maketitle
%------------------------------------------------------------------------------------%
\section{Introduction}

A topological index is a numeric value associated with a molecular structure and are used to correlates the physico-chemical properties of the molecular graph. In molecular graph, vertices corresponds to the atoms and edges corresponds to the bonds between them. Up to now, several topological indices have been defined and various mathematical properties and chemical applications are also investigated. Topological indices are used for studying quantitative structure-activity (QSAR) and structure-property (QSPR) relationship for predicting different properties of chemical compounds for modeling physicochemical, pharmacologic, toxicologic, biological and other properties of it and hence found different applications in chemistry, biochemistry and nanotechnology. The study of topological indices was introduced by American chemist Wiener for investigating boiling points of alkanes \cite{win47}. There exist various types of topological indices. Among them, the vertex degree based topological indices are most widely used and have different applications in chemical graph theory.

Throughout the paper, all graphs are finite, undirected and simple. Let $G$ be a connected graph with vertex set $V(G)$ and edge set $E(G)$ respectively. Let, $n$ and $m$ respectively denote the number of vertices and edges of $G$. The degree of a vertex is defined as the number of first neighbors of that vertex, so for any vertex ${v}\in V(G)$, if ${{d}_{G}}(v)$ denotes its degree and $N(v)$ denotes the set of vertices which are the neighbors of the vertex $v$, then ${{d}_{G}}(v)=|N(v)|$.

One of the oldest and well known topological indices is the first and second Zagreb indices, was first introduced by Gutman et al. in 1972 \cite{gutm72}, where they have examined the dependence of the total $\pi$-electron energy on molecular structure.
The first and second Zagreb indices of a graph are denoted by $M_1(G)$ and $M_2(G)$ and are respectively defined as

\[{{M}_{1}}(G)=\sum\limits_{v\in V(G)}{{{d}_{G}}{{(v)}^{2}}}=\sum\limits_{uv\in E(G)}{[{{d}_{G}}(u)+{{d}_{G}}(v)]}\] and \[{{M}_{2}}(G)=\sum\limits_{uv\in E(G)}{{{d}_{G}}(u){{d}_{G}}(v)}.\]

These  indices  are  extensively  studied  in  chemical and mathematical literature. Interested  readers  are referred to \cite{kha09,fath11,zho07,zho05,xu15,das15} for some recent results on the topic.

In 1984, Narumi and Katayama \cite{nar84} introduced a multiplicative graph invariant for representing the carbon skeleton of a saturated hydrocarbon, and named it as a ``simple topological index". Tomovic and Gutman \cite{tom01} later renamed this index as ``Narumi-Katayama index" or NK index and is denoted by $NK(G)$. The Narumi-Katayama index of a graph $G$ is defined as the product of degrees of all its vertices, that is

\[NK(G)=\prod\limits_{v\in V(G)}{{{{d}}_{G}}(v)}.\]

There are various mathematical and chemical study of Narumi-Katayama index, for details see \cite{gut12,kel10,hos13,hoss14,aza14}.

In 2010, Todeschini et al. \cite{tod10,tod10a} have introduced the multiplicative variants of additive graph invariants, which applied to the Zagreb indices would lead to the first and second Multiplicative Zagreb Indices. Thus the multiplicative Zagreb indices are defined as

\[\prod\nolimits_{1}{(G)}=\prod\limits_{uv\in E(G)}{{{d}_{G}}{{(u)}^{2}}}\] and \[\prod\nolimits_{2}{(G)}=\prod\limits_{uv\in E(G)}{{{d}_{G}}(u){{d}_{G}}(v)}.\]

The properties of these multiplicative Zagreb indices for trees were studied by Gutman \cite{gut11}.

Related to multiplicative version of ordinary first Zagreb index, Eliasi, Iranmanesh and Gutman \cite{eli12}, in 2012 introduced a new multiplicative graphical invariant and called multiplicative sum Zagreb index, which is defined as

\[\prod\nolimits_{1}{^{*}(G)}=\prod\limits_{uv\in E(G)}{\left[ {{d}_{G}}(u)+{{d}_{G}}(v) \right]}\]

These multiplicative Zagreb indices and multiplicative sum Zagreb index were subject to a large number of studies \cite{xu12,liu12,reti12,das13,aza15}.

Clearly, the Narumi-Katayama index is just the square root of the first multiplicative Zagreb index. In the following section, we proceed to introduce different properties of transformation and total transformation graphs and hence present the explicit expression of the Narumi-Katayama index of these transformation graphs in terms of some other graph invariants. First, we recall a well-known inequalities

\begin{lemma}
(A.M.-G.M. Inequality)
Let ${{x}_{1}},{{x}_{2}},....,{{x}_{n}}$be non-negative numbers, then
\[\frac{{{x}_{1}}+{{x}_{2}}+....+{{x}_{n}}}{n}\ge \sqrt[n]{{{x}_{1}}{{x}_{2}}....{{x}_{n}}}\]
with equality if and only if ${{x}_{1}}={{x}_{2}}=....={{x}_{n}}$.
\end{lemma}

\section{Background}

As different general and particular graphs of chemical interest can be obtained from simpler graphs via various graph transformations, hence it is important to study how different topological indices of such transformation are related to the corresponding topological indices of the original graph. Since, a transformation graphs converts the information from the original graph into new transformed structure, so if is it possible to find out the given graph from the transformed graph, then such operation may be used to figure out structural properties of the original graph considering the transformation graphs (for details see \cite{wu01}). Different graph transformation have been defined and studied in different mathematical literature, where the vertex set of the transformed graph is equal to $V(G)\cup E(G)$. Total graph is one of the best known type of such graphs. The total graph $T(G)$ of a graph $G$ is  the  graph  whose  vertex  set  is $V(G)\cup E(G)$ and any two vertices of $T(G)$ are adjacent if and only if they are either incident or adjacent in $G$ \cite{yan07, nd14a,nd15,nd15a}. Sampathkumar in \cite{sam1,sam2,sam3} introduced two other kinds of transformation graphs and were named as semitotal-point graph and semitotal-line graph. The semitotal-point graph of a graph $G$ is denoted by ${{T_1}}(G)$ and any two vertices $u,v\in V({{T}_{1}}(G))$ are adjacent if and only if either both $u$ and $v$ are adjacent vertices in $G$ or one is a vertex of $G$ and the other is an edge of $G$ incident to the former. So ${{T}_{1}}(G)$ has total $(m+n)$ vertices and $3m$ number of edges. The semitotal-line graph of a graph $G$ is denoted by ${{T_2}}(G)$ and any two vertices $u,v\in V({{T}_{2}}(G))$ are adjacent if and only if either both $u$ and $v$ are adjacent edges in $G$ or one is a vertex of $G$ and other is an edge of $G$ which is incident to it. So ${{T}_{1}}(G)$ has total $(m+n)$ vertices and $(\frac{1}{2}{{M}_{1}}(G)+m)$ number of edges.

The transformation graphs of $G$ has the vertex set $V(G)\cup E(G)$ and for any two vertices $u, v$ of the transformation graph, the associatively is $+$ if they are adjacent or incident in $G$  and  the associatively is $-$ if they are neither adjacent nor incident in $G$. In this paper we consider both the 2 and 3-permutation of the set {+,-}, that means transformation graphs of type ${{G}^{ab}}$ and ${{G}^{xyz}}$.

For the transformation graph ${{G}^{ab}}$, any two vertices $u,v\in V({{G}^{ab}})$,  $u$ and $v$ are adjacent in ${{G}^{ab}}$,  if and only if

(i) $u,v\in V(G)$, $u$, $v$ are adjacent in $G$ if $a= +$ and $u$, $v$ are not adjacent in $G$ if $a= -$.

(ii) $u,v\in E(G)$, $u$, $v$ are adjacent in $G$ if $b= +$ and $u$, $v$ are not adjacent in $G$ if $b= -$.

Since there are four distinct 2-permutation of ${+,-}$, so we can construct four different transformation of a given graph $G$ of this type. The degree of the vertices of these type of transformation graphs are given in the following proposition.

\begin{Proposition}

(i) ${{d}_{T(G)}}(u)=2{{d}_{G}}(u)$, if $u\in V(T(G))\cap V(G)$ and ${{d}_{T(G)}}(u)=2$, if $u\in V(T(G))\cap E(G)$.

(i) ${{d}_{{{G}^{++}}}}(u)=2{{d}_{G}}(u)$, if $u\in V({{G}^{++}})\cap V(G)$ and ${{d}_{{{G}^{++}}}}(u)=2$, if $u\in V({{G}^{++}})\cap E(G)$.

(ii) ${{d}_{{{G}^{+-}}}}(u)=m$, if $u\in V({{G}^{+-}})\cap V(G)$ and ${{d}_{{{G}^{+-}}}}(u)=n-2$, if $u\in V({{G}^{+-}})\cap E(G)$.

(iii) ${{d}_{{{G}^{-+}}}}(u)=n-1$, if $u\in V({{G}^{-+}})\cap V(G)$ and ${{d}_{{{G}^{-+}}}}(u)=2$, if $u\in V({{G}^{-+}})\cap E(G)$.

(iv) ${{d}_{{{G}^{--}}}}(u)=n+m-1-2{{d}_{G}}(u)$, if $u\in V({{G}^{--}})\cap V(G)$ and ${{d}_{{{G}^{--}}}}(u)=n-2$, if $u\in V({{G}^{--}})\cap E(G)$.
\end{Proposition}

Let  $x, y, z$ be three variables taking values either $+$ or $-$. Wu and Meng \cite{wu01} generalized the concept of total graphs to a total transformation graph ${{G}^{xyz}}$ with $x, y, z\epsilon \{-, + \}$. This type of transformation graphs of $G$ has the vertex set $V(G)\cup E(G)$ so that $|V({{G}^{xyz}})|=(n+m)$ and for any two vertices $u$, $v$ of the transformation graph, the associatively is $+$ if they are adjacent or incident in $G$  and  the associatively is $-$ if they are neither adjacent nor incident in $G$. Hence, for any two vertices $u,v\in V({{G}^{xyz}})$, $u$ and $v$ are adjacent in ${{G}^{xyz}}$, if and only if

(i) $u,v\in V(G)$, $u$, $v$ are adjacent in $G$ if $x = +$ and $u$, $v$ are not adjacent in $G$ if $x = -$.

(ii) $u,v\in E(G)$, $u$, $v$ are adjacent in $G$ if $y = +$ and $u$, $v$ are not adjacent in $G$ if $y= -$.

(iii) $u\in V(G)$ and $v\in E(G)$, $u$, $v$ are incident in $G$ if $z= +$ and $u$, $v$ are not incident in $G$ if $z= -$.

Clearly, there are eight distinct 3-permutation of ${+,-}$. Thus one can construct eight different transformation of a given graph $G$ of type ${{G}^{xyz}}$. From construction, it can be observed that the graphs ${{G}^{+++}}$ is isomorphic to the total graph $T(G)$. Also, from definition of this type of total transformation graphs, it is clear that, for a given graph $G$, ${{G}^{+++}}$ and ${{G}^{---}}$, ${{G}^{++-}}$ and ${{G}^{--+}}$, ${{G}^{-++}}$ and ${{G}^{+--}}$, ${{G}^{+-+}}$ and ${{G}^{-+-}}$ are the pairs of complementary graphs. For further study, interested reader are refer to \cite{xu08,yi09,wu01,zha02} for the study of basic properties of these transformation graphs where as different results of various topological indices such as Zagreb indices, F-index of these transformations are given in \cite{hos14,bas15,nd16}.

In this paper, we determine the Narumi-Katayama index of the transformation graphs and total transformation graphs of type ${{G}^{ab}}$ and ${{G}^{xyz}}$, that means both the 2 and 3-permutation of the set {+,-} in terms of different graph invariants and showed that how it changes under different transformation operations.

\section{Main Results}

In this section first we derive some explicit expressions of semitotal point graph and transformation graphs of type ${{G}^{ab}}$,  where  $a$, $b$ be the variables taking values either $+$ or $-$.

\begin{thm}
Let G be a graph of order n and size m. Then
\[NK({{T}_{1}}(G))={2^{(n+m)}}NK(G)\]
\end{thm}

\n\textit{Proof.} Using Proposition 1 we get the results from the following direct calculations.
\begin{eqnarray*}
NK({{T}_{1}}(G))&=&\prod\limits_{u\in V({{T}_{1}})}{{{d}_{{{T}_{1}}}}(u)}\\
               &=&\prod\limits_{u\in V({{T}_{1}})\cap V(G)}{{{d}_{{{T}_{1}}}}(u)}\prod\limits_{u\in V({{T}_{1}})\cap E(G)}{{{d}_{{{T}_{1}}}}(u)}\\
               &=&\prod\limits_{u\in V(G)}2{{{d}_{G}}(u)}\prod\limits_{uv\in E(G)}{2}\\
               &&={2^{(n+m)}}NK(G).
\end{eqnarray*}
Hence the result follows.  \qed

\begin{thm}
Let G be a graph of order n and size m. Then
\[NK({{G}^{++}})=NK({{T}_{2}}(G))=NK(G)\,\Pi _{1}^{*}(G).\]
\end{thm}

\n\textit{Proof.} From definition of Narumi-Katayama index, using proposition 1, we have
\begin{eqnarray*}
NK({{G}^{++}})&=&NK({{T}_{2}}(G))=\prod\limits_{u\in V({{G}^{++}})}{{{d}_{{{G}^{++}}}}(u)}\\
               &=&\prod\limits_{u\in V({{G}^{++}})\cap V(G)}{{{d}_{{{G}^{++}}}}(u)}\prod\limits_{u\in V({{G}^{++}})\cap E(G)}{{{d}_{{{G}^{++}}}}(u)}\\
               &=&\prod\limits_{u\in V(G)}{{{d}_{G}}(u)}\prod\limits_{uv\in E(G)}{[{{d}_{G}}(u)+{{d}_{G}}(v)]}\\
               &&=NK(G)\,\Pi _{1}^{*}(G).
\end{eqnarray*}
Hence the result.  \qed

\begin{thm}
Let G be a graph of order n and size m. Then
\[NK({{G}^{+-}})={{m}^{n}}.{{(n-2)}^{m}}.\]
\end{thm}

\n\textit{Proof.} Applying proposition 1 and from definition of Narumi-Katayama index we have
\begin{eqnarray*}
NK({{G}^{+-}})&=&\prod\limits_{u\in V({{G}^{+-}})}{{{d}_{{{G}^{+-}}}}(u)}\\
              &=&\prod\limits_{u\in V({{G}^{+-}})\cap V(G)}{{{d}_{{{G}^{+-}}}}(u)}\prod\limits_{u\in V({{G}^{+-}})\cap E(G)}{{{d}_{{{G}^{+-}}}}(u)}\\
              &=&\prod\limits_{u\in V(G)}{m}\prod\limits_{u\in E(G)}{(n-2)}\\
              &=&{{m}^{n}}.{{(n-2)}^{m}}.
\end{eqnarray*}
Hence we get the desired result.\qed

\begin{thm}
Let G be a graph of order n and size m. Then
\[NK({{G}^{-+}})={{2}^{m}}{{(n-1)}^{n}}.\]
\end{thm}
\n\textit{Proof.} From definition of Narumi-Katayama index and proposition 1 we similarly get
\begin{eqnarray*}
NK({{G}^{-+}})&=&\prod\limits_{u\in V({{G}^{-+}})}{{{d}_{{{G}^{-+}}}}(u)}\\
              &=&\prod\limits_{u\in V({{G}^{-+}})\cap V(G)}{{{d}_{{{G}^{-+}}}}(u)}\prod\limits_{u\in V({{G}^{-+}})\cap E(G)}{{{d}_{{{G}^{-+}}}}(u)}\\
              &=&\prod\limits_{u\in V(G)}{(n-1)}\prod\limits_{u\in E(G)}{2}\\
              &=&{{(n-1)}^{n}}{{.2}^{m}}.
\end{eqnarray*}
Hence the desired result follows. \qed
\begin{thm}
Let G be a graph of order n and size m. Then
\[NK({{G}^{--}})\le {{(n-2)}^{n}}{{\left[ (n+m-1)-\frac{4m}{n} \right]}^{n}}\]
with equality if and only if G be a regular graph.
\end{thm}

\n\textit{Proof.} Similarly, using proposition 1 and from definition of Narumi-Katayama index, we get
\begin{eqnarray*}
NK({{G}^{--}})&=&\prod\limits_{u\in V({{G}^{--}})}{{{d}_{{{G}^{--}}}}(u)}\\
              &=&\prod\limits_{u\in V({{G}^{--}})\cap V(G)}{{{d}_{{{G}^{--}}}}(u)}\prod\limits_{u\in V({{G}^{--}})\cap E(G)}{{{d}_{{{G}^{--}}}}(u)}\\
              &=&\prod\limits_{u\in V(G)}{[n+m-1-2{{d}_{G}}(u)]}\prod\limits_{u\in E(G)}{(n-2)}\\
\end{eqnarray*}
Now using lemma 1, we get
\begin{eqnarray*}
      \prod\limits_{u\in V(G)}{[n+m-1-2{{d}_{G}}(u)]}\le \sum\limits_{u\in V(G)}{[n+m-1-2{{d}_{G}}(u)]}={{\left[ (n+m-1)-\frac{4m}{n} \right]}^{n}}
\end{eqnarray*}
with equality if and only if $G$ be a regular graph. Hence the desired result follows from above.   \qed

\begin{cor}
If G  be a regular graph then
 \[NK({{G}^{--}})={{(n-2)}^{n}}{{\left[ (n+m-1)-\frac{4m}{n} \right]}^{n}}.\]
\end{cor}

Next we present the expressions of Narumi-Katayama index for total transformation graphs of two different classes of regular graphs namely cycle graph and complete graph with $n$ vertices.

\begin{ex}
(i) $NK({{C}_{n}}^{++})={{2}^{3n}}$.\\
(ii) $NK({{C}_{n}}^{+-})={{n}^{n}}{{(n-2)}^{n}}.$\\
(iii) $NK({{C}_{n}}^{-+})={{2}^{n}}{{(n-1)}^{n}}.$\\
(iv) $NK({{C}_{n}}^{--})={{(n-2)}^{n}}{{(2n-5)}^{n}}.$
\end{ex}

\begin{ex}
(i) $NK({{K}_{n}}^{++})={{2}^{\frac{n(n+1)}{2}}}{{(n-1)}^{n}}.$\\
(ii) $NK({{K}_{n}}^{+-})={{\left[ \frac{n(n-1)}{2} \right]}^{n}}{{(n-2)}^{\frac{n(n-1)}{2}}}.$\\
(iii) $NK({{K}_{n}}^{-+})={{2}^{\frac{n(n-1)}{2}}}{{(n-1)}^{n}}.$\\
(iv) $NK({{K}_{n}}^{--})={{(n-2)}^{\frac{n(n+1)}{2}}}{{\left[ \frac{(n-1)}{2} \right]}^{n}}.$\\
\end{ex}

Next we consider Narumi-Katayama index of the total transformation graphs of type ${{G}^{xyz}}$,  where  $x, y, z$ be three variables taking values either $+$ or $-$. Now, in the following first we calculate Narumi-Katayama index of the total transformation graphs of type ${{G}^{xyz}}$, where $x=y=z=+$.

\begin{thm}
Let G be a graph of order n and size m. Then
	\[NK({{G}^{+++}})={{2}^{n}}NK(G)\,\Pi _{1}^{*}(G).\]
\end{thm}

\n\textit{Proof.} From the construction of ${{G}^{+++}}$ we have, for  $v\in V({{G}^{+++}})\cap V(G)$, ${{d}_{{{G}^{+++}}}}(v)=2{{d}_{G}}(v)$  and  for $v\in V({{G}^{+++}})\cap E(G)$, ${{d}_{{{G}^{+++}}}}(v)={{d}_{G}}(u)+{{d}_{G}}(v)$.
\begin{eqnarray*}
NK({{G}^{+++}})&=&\prod\limits_{v\in V({{G}^{+++}})}{{{d}_{{{G}^{+++}}}}(v)}\\
&=&\prod\limits_{v\in V({{G}^{+++}})\cap V(G)}{{{d}_{{{G}^{+++}}}}(v)}\prod\limits_{v\in V({{G}^{+++}})\cap E(G)}{{{d}_{{{G}^{+++}}}}(v)}\\
&=&\prod\limits_{v\in V(G)}{2{{d}_{G}}(u)}\prod\limits_{uv\in E(G)}{[{{d}_{G}}(u)+{{d}_{G}}(v)]}\\
&=&{{2}^{n}}NK(G)\,\Pi _{1}^{*}(G).
\end{eqnarray*}
Hence the desired result follows. \qed

Next, in the following we calculate Narumi-Katayama index of the complement of total transformation graphs of above type, that is of ${{G}^{xyz}}$, where $x=y=z=-$.

\begin{thm}
Let G be a graph of order n and size m. Then
	\[NK({{G}^{---}})\le {{\left[ (m+n-1)-\frac{4m}{n} \right]}^{n}}{{\left[ (m+n-1)-\frac{1}{m}{{M}_{1}}(G) \right]}^{m}}\]
with equality if and omly if G be a regular graph.
\end{thm}

\n\textit{Proof.} From the construction of ${{G}^{---}}$ we have, for  $v\in V({{G}^{---}})\cap V(G)$, ${{d}_{{{G}^{---}}}}(v)=m+n-1-2{{d}_{G}}(v)$  and  for $v\in V({{G}^{---}})\cap E(G)$, ${{d}_{{{G}^{---}}}}(v)=m+n-1-({{d}_{G}}(u)+{{d}_{G}}(v))$.

\begin{eqnarray*}
NK({{G}^{---}})&=&\prod\limits_{v\in V({{G}^{---}})}{{{d}_{{{G}^{---}}}}(v)}\\
               &=&\prod\limits_{v\in V({{G}^{---}})\cap V(G)}{{{d}_{{{G}^{---}}}}(v)}\prod\limits_{v\in V({{G}^{---}})\cap E(G)}{{{d}_{{{G}^{---}}}}(v)}\\
               &=&\prod\limits_{v\in V(G)}{\left[ m+n-1-2{{d}_{G}}(v) \right]}\prod\limits_{uv\in E(G)}{[(m+n-1)-({{d}_{G}}(u)+{{d}_{G}}(v))]}.
\end{eqnarray*}
Now using lemma 1, we get
\[\prod\limits_{v\in V(G)}{[n+m-1-2{{d}_{G}}(u)]}\le \sum\limits_{v\in V(G)}{[n+m-1-2{{d}_{G}}(u)]}={{\left[ (n+m-1)-\frac{4m}{n} \right]}^{n}}\]
and
\begin{eqnarray*}
\prod\limits_{uv\in E(G)}{[(n+m-1)-({{d}_{G}}(u)+{{d}_{G}}(v))]}&\le& \sum\limits_{uv\in E(G)}{[(n+m-1)-({{d}_{G}}(u)+{{d}_{G}}(v))]}\\
&=&{{\left[ (n+m-1)-\frac{{{M}_{1}}(G)}{m} \right]}^{m}}
\end{eqnarray*}
with equality if and only if $G$ be a regular graph. \qed

\begin{cor}
Let G be a r-regular graph of order n and size m. Then
\[NK({{G}^{---}})={{(m+n-2r-1)}^{m+n}}.\]
\end{cor}

Next, In the following, we calculate Narumi-Katayama index of the total transformation graphs of type ${{G}^{xyz}}$,  where  $x=y=+$ and $z=-$, that is of ${G}^{++-}$.

\begin{thm}
Let G be a graph of order n and size m. Then
\[NK({{G}^{++-}})\le {{m}^{n}}{{\left[ \frac{1}{m}{{M}_{1}}(G)+(n-4) \right]}^{m}}\]
with equality if and only if G be a regular graph.
\end{thm}
\n\textit{Proof.} We have, for  $v\in V({{G}^{++-}})\cap V(G)$, ${{d}_{{{G}^{++-}}}}(v)=m$  and  for $v\in V({{G}^{++-}})\cap E(G)$${{d}_{{{G}^{++-}}}}(v)={{d}_{G}}(u)+{{d}_{G}}(v)+n-4$.  Thus from (1) we have
\begin{eqnarray*}
NK({{G}^{++-}})&=&\prod\limits_{v\in V({{G}^{++-}})}{{{d}_{{{G}^{++-}}}}(v)}\\
               &=&\prod\limits_{v\in V({{G}^{++-}})\cap V(G)}{{{d}_{{{G}^{++-}}}}(v)}\prod\limits_{v\in V({{G}^{++-}})\cap E(G)}{{{d}_{{{G}^{++-}}}}(v)}\\
               &=&\prod\limits_{v\in V(G)}{m}\prod\limits_{uv\in E(G)}{[{{d}_{G}}(u)+{{d}_{G}}(v)+(n-4)]}\\
               &=&{{m}^{n}}\prod\limits_{uv\in E(G)}{[{{d}_{G}}(u)+{{d}_{G}}(v)+(n-4)]}.
\end{eqnarray*}
Now using lemma 1, we get
\[\prod\limits_{uv\in E(G)}{[{{d}_{G}}(u)+{{d}_{G}}(v)+(n-4)]}\le {{\left[ \frac{1}{m}{{M}_{1}}(G)+(n-4) \right]}^{m}}\]
with equality if and only if G be a regular graph. Hence the desired result follows.   \qed

\begin{cor}
Let G be a r-regular graph of order n and size m. Then
\[NK({{G}^{++-}})={{m}^{n}}{{(2r+n-4)}^{m}}.\]
\end{cor}

In the following, next we calculate Narumi-Katayama index of the complement of total transformation graphs of ${G}^{++-}$, that is of ${{G}^{--+}}$.

\begin{thm}
Let G be a graph of order n and size m. Then
\[NK({{G}^{--+}})\le {{(n-1)}^{n}}{{\left[ m+3-\frac{1}{m}{{M}_{1}}(G) \right]}^{m}}\]
with equality if and only if G be a regular graph.
\end{thm}
\n\textit{Proof.} We have,  for  $u\in V({{G}^{--+}})\cap V(G)$, ${{d}_{{{G}^{--+}}}}(u)=n-1$ and  for $u\in V({{G}^{--+}})\cap E(G)$${{d}_{{{G}^{--+}}}}(u)=(m+3)-({{d}_{G}}(u)+{{d}_{G}}(v))$.   Thus from (1) we have
\begin{eqnarray*}
NK({{G}^{--+}})&=&\prod\limits_{v\in V({{G}^{--+}})}{{{d}_{{{G}^{--+}}}}(v)}\\
               &=&\prod\limits_{v\in V({{G}^{--+}})\cap V(G)}{{{d}_{{{G}^{--+}}}}(v)}\prod\limits_{v\in V({{G}^{--+}})\cap E(G)}{{{d}_{{{G}^{--+}}}}(v)}\\
               &=&\prod\limits_{v\in V(G)}{(n-1)}\prod\limits_{uv\in E(G)}{[(m+3)-({{d}_{G}}(u)+{{d}_{G}}(v))]}\\
               &=&{{(n-1)}^{n}}\prod\limits_{uv\in E(G)}{[(m+3)-({{d}_{G}}(u)+{{d}_{G}}(v))]}.
\end{eqnarray*}
Again,  using lemma 1, we get
\[\prod\limits_{uv\in E(G)}{[(m+3)-({{d}_{G}}(u)+{{d}_{G}}(v))]}\le {{\left[ m+3-\frac{1}{m}{{M}_{1}}(G) \right]}^{m}}\]
with equality if and only if G be a regular graph.  Hence the desired result follows from above. \qed

\begin{cor}
Let G be a r-regular graph of order n and size m. Then
\[NK({{G}^{--+}})={{(n-1)}^{n}}{{(m+3-2r)}^{m}}.\]
\end{cor}

Now, in the following, we consider F-index of the total transformation graphs of type ${{G}^{xyz}}$,  where  $x=-$ and $y=z=+$, that is of ${G}^{-++}$.

\begin{thm}
Let G be a graph of order n and size m. Then
\[NK({{G}^{-++}})={{(n-1)}^{n}}\,\Pi _{1}^{*}(G).\]
\end{thm}

\n\textit{Proof.} From definition we have, for  $v\in V({{G}^{-++}})\cap V(G)$, ${{d}_{{{G}^{-++}}}}(u)=(n-1)$  and  for $v\in V({{G}^{-++}})\cap E(G)$${{d}_{{{G}^{-++}}}}(u)={{d}_{G}}(u)+{{d}_{G}}(v)$.
\begin{eqnarray*}
NK({{G}^{-++}})&=&\prod\limits_{v\in V({{G}^{-++}})}{{{d}_{{{G}^{-++}}}}(v)}\\
               &=&\prod\limits_{v\in V({{G}^{-++}})\cap V(G)}{{{d}_{{{G}^{-++}}}}(v)}\prod\limits_{v\in V({{G}^{-++}})\cap E(G)}{{{d}_{{{G}^{-++}}}}(v)}\\
               &=&\prod\limits_{v\in V(G)}{(n-1)}\prod\limits_{uv\in E(G)}{[{{d}_{G}}(u)+{{d}_{G}}(v)]}\\
               &=&{{(n-1)}^{n}}\,\Pi _{1}^{*}(G).
\end{eqnarray*}
from where the desired result follows. \qed

Again, similarly we now consider the Narumi-Katayama index of the complement of the last type of graph ${G}^{-++}$, that is of ${G}^{+--}$.

\begin{thm}
Let G be a graph of order n and size m. Then
\[NK({{G}^{+--}})\le {{m}^{n}}{{\left[ (m+n-1)-\frac{1}{m}{{M}_{1}}(G) \right]}^{m}}\]
with equality if and only if G be a regular graph.
\end{thm}

\n\textit{Proof.} From definition, we have for  $v\in V({{G}^{+--}})\cap V(G)$, ${{d}_{{{G}^{+--}}}}(u)=m$  and  for $v\in V({{G}^{+--}})\cap E(G)$,${{d}_{{{G}^{+--}}}}(u)=(m+n-1)-({{d}_{G}}(u)+{{d}_{G}}(v))$.
\begin{eqnarray*}
NK({{G}^{+--}})&=&\prod\limits_{v\in V({{G}^{+--}})}{{{d}_{{{G}^{+--}}}}(v)}\\
               &=&\prod\limits_{v\in V({{G}^{+--}})\cap V(G)}{{{d}_{{{G}^{+--}}}}(v)}\prod\limits_{v\in V({{G}^{+--}})\cap E(G)}{{{d}_{{{G}^{+--}}}}(v)}\\
               &=&\prod\limits_{v\in V(G)}{m}\prod\limits_{uv\in E(G)}{[(m+n-1)-({{d}_{G}}(u)+{{d}_{G}}(v))]}\\
               &=&{{m}^{n}}\,\prod\limits_{uv\in E(G)}{[(m+n-1)-({{d}_{G}}(u)+{{d}_{G}}(v))]}.
\end{eqnarray*}
Again,  using lemma 1, we get
\[\prod\limits_{uv\in E(G)}{[(m+n-1)-({{d}_{G}}(u)+{{d}_{G}}(v))]}\le {{\left[ (m+n-1)-\frac{1}{m}{{M}_{1}}(G) \right]}^{m}}\]
with equality if and only if G be a regular graph.  Hence the desired result follows from above. \qed

\begin{cor}
Let G be a r-regular graph of order n and size m. Then
\[NK({{G}^{+--}})={{m}^{n}}{{(m+n-2r-1)}^{m}}.\]
\end{cor}

In the following, now we find Narumi-Katayama index of the total transformation graph of type ${G}^{+-+}$.

\begin{thm}
Let G be a graph of order n and size m. Then
\[NK({{G}^{+-+}})\le {{2}^{n}}NK(G){{\left[ (m+3)-\frac{1}{m}{{M}_{1}}(G) \right]}^{m}}\]
with equality if and only if G be a regular graph.
\end{thm}

\n\textit{Proof.} From definition we have, for  $v\in V({{G}^{+-+}})\cap V(G)$, ${{d}_{{{G}^{+-+}}}}(v)=2{{d}_{G}}(v)$  and  for $v\in V({{G}^{+-+}})\cap E(G)$${{d}_{{{G}^{+-+}}}}(v)=m+3-({{d}_{G}}(u)+{{d}_{G}}(v))$.
\begin{eqnarray*}
NK({{G}^{+-+}})&=&\prod\limits_{v\in V({{G}^{+-+}})}{{{d}_{{{G}^{+-+}}}}(v)}\\
               &=&\prod\limits_{v\in V({{G}^{+-+}})\cap V(G)}{{{d}_{{{G}^{+-+}}}}(v)}\prod\limits_{v\in V({{G}^{+-+}})\cap E(G)}{{{d}_{{{G}^{+-+}}}}(v)}\\
               &=&\prod\limits_{v\in V(G)}{2{{d}_{G}}(v)}\prod\limits_{uv\in E(G)}{[(m+3)-({{d}_{G}}(u)+{{d}_{G}}(v))]}\\
               &=&{{2}^{n}}NK(G)\prod\limits_{uv\in E(G)}{[(m+3)-({{d}_{G}}(u)+{{d}_{G}}(v))]}.
\end{eqnarray*}
Again,  using lemma 1, we get
                     \[\prod\limits_{uv\in E(G)}{[(m+3)-({{d}_{G}}(u)+{{d}_{G}}(v))]}\le {{\left[ (m+3)-\frac{1}{m}{{M}_{1}}(G) \right]}^{m}}\]
with equality if and only if $G$ be a regular graph. \qed

\begin{cor}
Let G be a r-regular graph of order n and size m. Then
\[NK({{G}^{+-+}})={{2}^{n}}{{r}^{n}}{{(m+3-2r)}^{m}}.\]
\end{cor}

Finally, in the following, we determine Narumi-Katayama index of the total transformation graph of complement of ${G}^{+-+}$, that is ${G}^{-+-}$.

\begin{thm}
Let G be a graph of order n and size m. Then
\[NK({{G}^{-+-}})\le {{\left[ m+n-1-\frac{4m}{n} \right]}^{n}}{{\left[ (n-4)+\frac{1}{m}{{M}_{1}}(G) \right]}^{m}}\]
with equality if and only if G be a regular graph.
\end{thm}

\n\textit{Proof.} From the costruction of ${{G}^{-+-}}$, we have,  for  $v\in V({{G}^{-+-}})\cap V(G)$, ${{d}_{{{G}^{-+-}}}}(u)=(m+n-1)-2{{d}_{G}}(v)$  and  for $v\in V({{G}^{-+-}})\cap E(G)$,${{d}_{{{G}^{-+-}}}}(u)=(n-4)+({{d}_{G}}(u)+{{d}_{G}}(v))$. Hence we have
\begin{eqnarray*}
NK({{G}^{-+-}})&=&\prod\limits_{v\in V({{G}^{-+-}})}{{{d}_{{{G}^{-+-}}}}(v)}\\
               &=&\prod\limits_{v\in V({{G}^{-+-}})\cap V(G)}{{{d}_{{{G}^{-+-}}}}(v)}\prod\limits_{v\in V({{G}^{-+-}})\cap E(G)}{{{d}_{{{G}^{-+-}}}}(v)}\\
               &=&\prod\limits_{v\in V(G)}{(m+n-1-2{{d}_{G}}(v))}\prod\limits_{uv\in E(G)}{[(n-4)+({{d}_{G}}(u)+{{d}_{G}}(v))]}.
\end{eqnarray*}
Again,  using lemma 1, we get
\[\prod\limits_{v\in V(G)}{(m+n-1-2{{d}_{G}}(v))}\le {{\left[ m+n-1-\frac{4m}{n} \right]}^{n}}\]
and
\[\prod\limits_{uv\in E(G)}{[(n-4)+({{d}_{G}}(u)+{{d}_{G}}(v))]}\le {{\left[ (n-4)+\frac{1}{m}{{M}_{1}}(G) \right]}^{m}}\]
with equality if and only if $G$ be a regular graph.  Hence the desired result follows. \qed

\begin{cor}
Let G be a r-regular graph of order n and size m. Then
\[NK({{G}^{-+-}})={{(m+n-2r-1)}^{n}}{{(n+2r-4)}^{m}}.\]
\end{cor}

Next we present the expressions of Narumi-Katayama index for total transformation graphs of two different classes of regular graphs namely ${{C}_{n}}$ and ${{K}_{n}}$.

\begin{ex}
Consider the cycle ${{C}_{n}}$ with n vertices.  Since, its every vertex is of degree 2, so the Narumi Katayama index of different total transformation graphs of ${{C}_{n}}$ are given by

(i) $NK({{C}_{n}}^{+++})={{4}^{2n}}.$

(ii) $NK({{C}_{n}}^{---})={{(2n-5)}^{2n}}.$

(iii) $NK({{C}_{n}}^{++-})={{n}^{2n}}.$

(iv) $NK({{C}_{n}}^{--+})={{(n-1)}^{2n}}.$

(v) $NK({{C}_{n}}^{+--})=NK({{C}_{n}}^{-+-})={{n}^{n}}{{(2n-5)}^{n}}.$

(vi) $NK({{C}_{n}}^{+-+})=NK({{C}_{n}}^{-++})={{4}^{n}}{{(n-1)}^{n}}.$
\end{ex}

\begin{ex}
Let ${{K}_{n}}$ be the complete graph on n vertices. All vertices of ${{K}_{n}}$ have degree  (n - 1) and so Narumi Katayama index of different total transformation graphs of ${{K}_{n}}$ are given by

(i) $NK({{K}_{n}}^{+++})={{4}^{n}}{{(n-1)}^{2n}}.$

(ii) $NK({{K}_{n}}^{---})={{\left[ \frac{(n-1)(n-2)}{2} \right]}^{\frac{n(n+1)}{2}}}.$

(iii) $NK({{K}_{n}}^{++-})={{\left[ \frac{n(n-1)}{2} \right]}^{n}}{{(3n-8)}^{m}}.$

(iv) $NK({{K}_{n}}^{--+})={{(n-1)}^{n}}{{\left[ \frac{n(n-1)}{2}-2n+1 \right]}^{m}}.$

(v) $NK({{K}_{n}}^{-++})={{2}^{\frac{n(n-1)}{2}}}{{(n-1)}^{\frac{n(n+1)}{2}}}.$

(vi) $NK({{K}_{n}}^{+--})={{\left[ \frac{n(n-1)}{2} \right]}^{n}}{{\left[ \frac{(n-1)(n-2)}{2} \right]}^{\frac{n(n-1)}{2}}}.$

(vii) $NK({{K}_{n}}^{+-+})={{2}^{n}}{{(n-1)}^{n}}{{\left[ \frac{n(n-1)}{2}-2n+5 \right]}^{\frac{n(n-1)}{2}}}.$

(viii) $NK(K_{n}^{-+-})={{\left[ \frac{(n-1)(n+6)}{2} \right]}^{n}}{{\left\{ 3(n-2) \right\}}^{\frac{n(n-1)}{2}}}.$

\end{ex}

\section{Conclusion}

In this paper, we determine the Narumi-Katayama index of the total transformation graphs in terms of different graph invariants and showed that how it changes under different transformations. Since, the Narumi-Katayama index is the square root of the  first multiplicative Zagreb index, so the expressions for first multiplicative Zagreb index of total transformation graphs can easily be obtained from the above derived expressions for Narumi-Katayama index. For further studies different other topological indices  of some new transformation graphs can be obtained to understand the underlying topology.

\end{document}